\documentclass[aps,prl,reprint,longbibliography,superscriptaddress]{revtex4-2}

 \usepackage{amsmath,bm}
 \usepackage{mathrsfs}
 \usepackage{amsfonts}
 \usepackage{graphicx}
 \usepackage{setspace} 
 \usepackage{graphicx}
 \usepackage{epstopdf}
 \usepackage{dcolumn}
 \usepackage{amsmath}
 \usepackage{epsfig}
 \usepackage{indentfirst}
 \usepackage{psfrag}
 \usepackage{subfigure}
 \usepackage{amssymb}
 \usepackage{color}
 \usepackage{units} 

 \usepackage{graphicx}
 \usepackage{dcolumn}
 \usepackage{bm}
 \usepackage{natbib}

\usepackage{physics}
\usepackage{dcolumn}
\usepackage{bm}
\usepackage{natbib}
\usepackage[backref=none,bookmarksnumbered=true,bookmarks=true,bookmarksopen=true,colorlinks=true,
citecolor=blue,linkcolor=blue,anchorcolor=green,urlcolor=blue,unicode=false]{hyperref}

\usepackage{ulem}[normalem] 

\makeatletter
\newcommand\colorsout[1]{\bgroup \markoverwith{\textcolor{#1}{\rule[0.5ex]{2pt}{0.4pt}}}\ULon}

\makeatother
\begin{document}

\title{Signatures of Kondo-lattice behavior in the two-dimensional ferromagnet Fe$_3$GeTe$_2$}

\author{Carmen Rubio-Verd\'u }
  \affiliation{\mbox{Fachbereich Physik, Freie Universit\"at Berlin, 14195 Berlin, Germany}}
  \affiliation{{ICFO - Institut de Ciencies Fotoniques, The Barcelona Institute of Science and Technology, Castelldefels (Barcelona) 08860, Spain}}
  \email{carmen.rubio@icfo.eu}
\author{Katharina J. Franke}
\affiliation{\mbox{Fachbereich Physik, Freie Universit\"at Berlin, 14195 Berlin, Germany}}

\begin{abstract}
Fe$_2$GeTe$_3$ is a paradigmatic van der Waals magnet with the full microscopic picture of magnetic interactions still under debate. Here, we use scanning tunneling microscopy and spectroscopy at 1.1\,K to map out the low-energy physics on the surface. In agreement with previous works, we observe a spatially varying Fano lineshape that has been ascribed to a Kondo resonance. In addition, we identify a small gap of $\approx$\,5\,meV width on top of the larger-energy Fano resonance. We suggest that this gap is the signature of a coherent Kondo lattice that has been proposed to originate from hybridization of a strongly dispersing and a flat band both derived from $d$ electrons of the Fe atoms in the lattice. Adding magnetic adatoms on the surface hardly influences the magnetic signatures of the substrate, indicating the robustness of the intralayer Kondo correlations against magnetic adsorbates.  

	\end{abstract}

    \maketitle
The interplay between localized magnetic moments and itinerant conduction electrons in strongly correlated materials has long been a subject of intense research \cite{Stewart1984}. One fascinating manifestation of this interaction is the Kondo lattice, where the coupling between localized spins and conduction electrons leads to a complex interplay between Kondo screening and magnetic ordering \cite{Doniach1977, Hewson1997, Dzero2016}. Kondo-lattice systems have been extensively studied in bulk intermetallic compounds containing rare earth atoms. The localized magnetic moments are provided by $f$ electrons that are exchange coupled to the itinerant conduction electrons, leading to Kondo screening and the formation of a Kondo lattice \cite{Schmidt2010, Ernst2011, Aynajian2012}. Contrary, in $d$ electron systems, heavy-fermion behavior is typically not observed owing to the absence of strongly localized magnetic moments. However, hybrid systems, in which localized electrons originate from an adsorbate structure containing transition metal atoms, coupled to an electron bath in the substrate, have exhibited signatures of extended Kondo correlations \cite{Lagares2019, Vano2021, Wan2023}.

 With the advent of two-dimensional (2D) materials, new opportunities for exploring competing interactions and correlation effects entered the stage. In reduced dimension, quantum confinement and enhanced electronic correlations add versatility and novel phenomenology, including superconductivity, Mott insulating behavior, charge-density modulations \cite{Novoselov2016}, and in particular also ferromagnetism \cite{Gibertini2019} and Kondo-lattice behavior \cite{Zhang2018}.

It was only in the recent years that the first 2D magnets were isolated. One of the first was CrI$_3$, which shows ferromagnetic or antiferromagnetic behavior depending on the number of layers \cite{Huang2017}, followed by ferromagnetic Cr$_2$Ge$_2$Te$_6$, which loses its magnetic order in the monolayer limit \cite{Gong2017}.  Fe$_3$GeTe$_2$ is a ferromagnet with a high T$_C$ of about 220 K \cite{Deiseroth2006, Chen2013, May2016} even at monolayer thickness \cite{Fei2018}, which makes it particularly attractive for spintronic applications. 
However, closer examination of this material reveals that its magnetic behavior is far more complex than that of a simple ferromagnet. A large set of experimental observations revealed seemingly contradictory results. Signatures of both itinerant ferromagnetism and large effective electron mass indicating the presence of localized magnetic moments were observed, launching also intense theoretical efforts to resolve these apparent contradictions \cite{Xu2020, Yang2022, Zhao2021_holes, Bao2022}. The co-existence of itinerant and local magnetism has been explained by either the multi-orbital character of the Fe atoms \cite{Bai2022} or two distinct Fe sites exhibiting different Mott and Hund correlation strengths \cite{Zhu2016, Corasantini2020, Kim2022, Bao2022, Xu2024}. 

The role of orbital- and site-differentiated magnetic contributions are also central to the debate about heavy-fermion behavior and Kondo physics in this system. In particular, the observation of a Kondo lattice would otherwise be incompatible with a ferromagnetic ground state. However, despite indications of Kondo physics, an unambiguous signature of a Kondo lattice remains elusive.
Scanning tunneling microscopy (STM) experiments revealed a Fano-shaped dip at the Fermi level, whose line profile varies across the surface. This spectral fingerprint was assigned to a Kondo resonance. Given the omnipresent Fano resonance (albeit with variations in shape) across the surface combined with complementary indications of heavy-fermion behavior, it was suggested that the Kondo impurities form indeed a Kondo lattice \cite{Zhang2018, Zhao2021_holes}. However, this interpretation was challenged by more recent STM experiments, which suggested that the observed features in the same energy range may instead arise from phonon-magnon excitations \cite{Bansal2023} or from the suppression of the density of states by disorder in the correlated system \cite{Mathimalar2025}. This discussion echoes the earlier apparent controversies in the bulk material, where two-site and multi-orbital models were introduced to capture the contradicting observations. A crucial step toward resolving the Kondo-lattice interpretation would be the detection of a hybridization gap superimposed on the Fano lineshape as a clear signature of Kondo-lattice formation \cite{Schmidt2010, Hamidian2011, Aynajian2012}.

\begin{figure}[t!]
    \includegraphics[width=0.95\columnwidth]{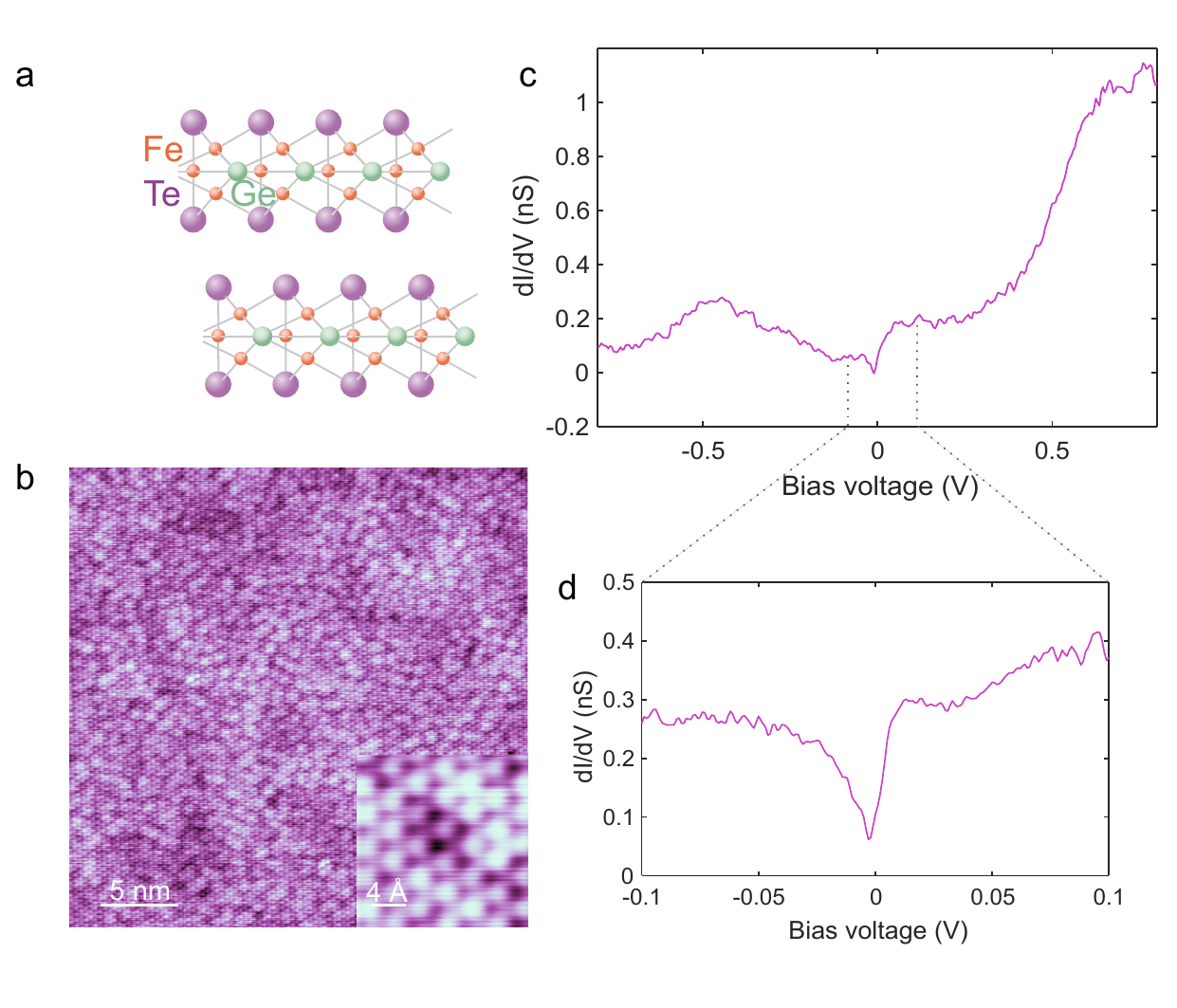}
	\caption{\label{Fig1} 
	Overview of Fe$_3$GeTe$_2$. (a) Crystal structure of Fe$_3$GeTe$_2$. The atomic species are color coded as shown. (b) STM topography of the Fe$_3$GeTe$_2$ surface. Inset: close-up view on some defects in an atomically-resolved STM image (V$_\mathrm{S}$ = 200 mV, I$_\mathrm{t}$ = 150 pA).
	(c)  d$I$/d$V$ spectrum obtained on the Fe$_3$GeTe$_2$ surface. (V$_\mathrm{S}$ = 800 mV, I$_\mathrm{t}$ = 300 pA, V$_\mathrm{mod}$ = 5 mV). (d) Close-up view on d$I$/d$V$ spectrum around the Fermi energy with an asymmetric dip. (V$_\mathrm{S}$ = 100 mV, I$_\mathrm{t}$ = 300 pA, V$_\mathrm{mod}$ = 1 mV).
	Analysis of STM topographs and spectroscopy data was performed with the WSxM \cite{wsxm} and SpectraFox \cite{Spectrafox} software packages.}
\end{figure}

In this work, we contribute to the ongoing debate by presenting high-resolution scanning tunneling microscopy and spectroscopy measurements of low-energy excitations on the surface of a bulk Fe$_3$GeTe$_2$ sample at 1.1\,K. Our spectra reproduce the previously observed Fano lineshape while also revealing an additional small gap of $\approx$\,5\,meV width superimposed on the broader Fano feature. This small gap rapidly disappears with increasing temperature, whereas the overall Fano lineshape remains intact. We propose that this gap represents the characteristic hybridization gap, providing further indication for the Kondo-lattice and heavy-fermion behavior in Fe$_3$GeTe$_2$. 
Additionally, we employ magnetic adatoms as probes to assess the balance of competing interactions in the low-temperature many-body state. Our results show that individual iron (Fe) and manganese (Mn) adatoms mainly affect the asymmetry parameter of the Fano lineshape owing to additional co-tunneling paths but it does not alter the other Kondo parameters. This indicates that the many-body phase coherence in the substrate remains robust against small magnetic perturbations.

The crystal structure of Fe$_3$GeTe$_2$ (shown in Fig.\,\ref{Fig1}a) consists of Te-terminated van-der-Waals-stacked layers. The Fe atoms are located in two different Wyckhoff positions. The central plane is formed by Fe-Ge, with the Fe atoms typically referred to as Fe2, sandwiched between planes of hexagonal Fe arrangements, where the Fe atoms are called Fe1. The Fe atoms not only differ in their crystal-field symmetries, but also in their contribution to the electronic structure and magnetic properties. Indeed, the seemingly contradictory magnetic behavior resembling Stoner-type and Heisenberg-type magnetism has been discussed to be rooted in the different types of Fe atoms, likely arising from different Hunds coupling strengths \cite{Xu2020}. 

Large-scale STM images taken after cleaving the bulk crystal in ultra-high vacuum reveal atomically flat terraces of the terminating Te layer (Figure\,\ref{Fig1}b). Alongside the atomic corrugation, a non-periodic modulation of the apparent height is observed. A magnified view in the inset of Fig.\,\ref{Fig1}b highlights one of such depressions, approximately 60\,pm in apparent depth. These features have been attributed to randomly distributed Fe2 vacancies in the near-surface layers, which locally modify the density of states \cite{Zhao2021_holes}. Notably, despite these inhomogeneities, long-range ferromagnetism remains intact, persisting over domains spanning several hundred nanometers \cite{Nguyen2018, Trainer2022, Yang2022}. In the following, we examine if this nanoscale disorder influences the Kondo-lattice behavior.

A differential conductance (d$I$/d$V$) spectrum (Fig.\,\ref{Fig1}c) shows a broad peak centered around -0.5\,V and a strong increase in conductance above 0.5\,V. Additionally, a narrow dip appears at the Fermi level, as highlighted in the close-up view in Fig.\,\ref{Fig1}d. The dip features a Fano lineshape of about 10\,meV width. This signature has been observed previously and was attributed to a Kondo resonance \cite{Zhang2018, Zhao2021_holes}.

\begin{figure*}[t!]
    \includegraphics[width=0.9\textwidth]{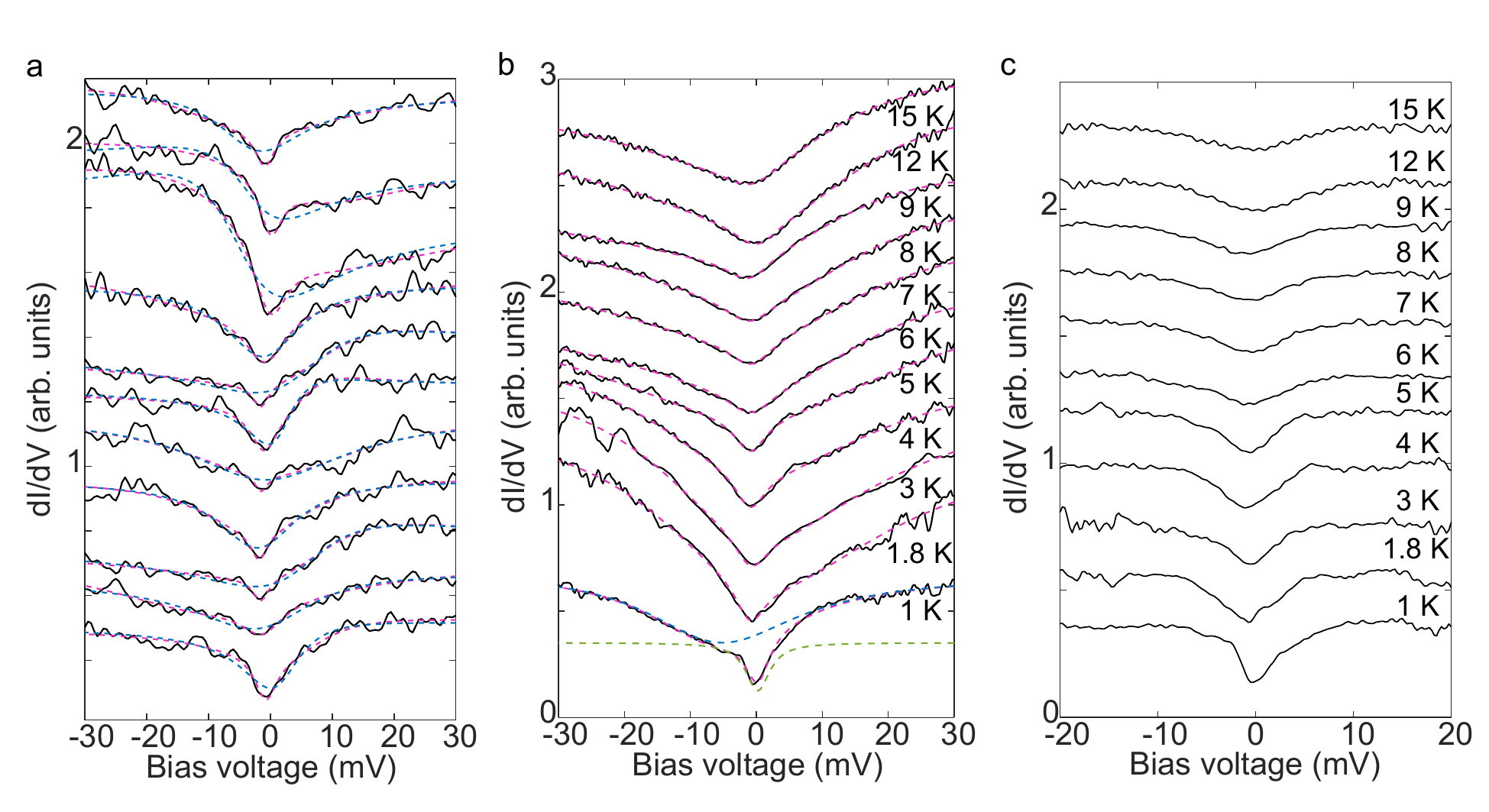}
	\caption{\label{Fig2} 
	Low-energy spectra. (a)  d$I$/d$V$ spectra obtained on multiple locations of the Fe$_2$GeTe$_3$ surface. The spectra are fit using two models: one with a single Fano lineshape, and another combining a Fano lineshape with a gap function, as described in the text (solid black: raw data, pink dashed: Fano and gap fit, blue dashed: Fano fit). (V$_\mathrm{S}$ = 40 mV, I$_\mathrm{t}$ = 300 pA, V$_\mathrm{mod}$ = 0.5 mV). (b) Temperature dependence of spectra on the Fe$_2$GeTe$_3$ surface. (V$_\mathrm{S}$ = 35 mV, I$_\mathrm{t}$ = 500 pA, V$_\mathrm{mod}$ = 0.25 mV). The blue and green dashed lines are the Fano and gap contributions to the curve d$I$/d$V$ curve at lowest temperature, respectively. The pink dashed line is the fit to the Fano and gap mixed function. (c) Data after subtracting the broad Fano lineshape at each temperature. 
 }
\end{figure*}

To gain more detailed information, we recorded high-resolution spectra in this energy range at randomly selected locations on the surface. A series of these spectra is presented in Fig.\,\ref{Fig2}a, all of which exhibit a dip feature around the Fermi level.
However, the exact lineshape varies across different locations. To quantify these variations and gain insight into the origin of the excitation spectrum, we first fit the data using a Fano function to the conductance $\sigma = A \cdot \frac{(\epsilon + q)^2}{(1+\epsilon ^2)} + \sigma_0$  with $\epsilon = (eV-E_\mathrm{R})/(\Gamma_\mathrm{R}/2)$. This lineshape is expected for tunneling into isolated Kondo impurities on surfaces as well as a Kondo lattice \cite{Yang2009, Figgins2011} and includes the effect of interfering tunneling paths. Here $q$ is the asymmetry parameter which accounts for the ratio between the tunneling paths; $E_\mathrm{R}$ is the energy of the resonance and $\Gamma_\mathrm{R}$ corresponds to its full-width. The amplitude of the resonance is accounted by the parameter $A$ and $\sigma_0$ represents the background conductance.
In analyzing the best fits, we notice that the spectra cannot be completely reproduced by a single Fano function. Instead, we find an additional narrow dip at the Fermi level. To fully capture the lineshape, we thus add a a Lorentzian dip ($\sigma_{\mathrm{gap}} = A_{\mathrm{gap}} \cdot \frac{\Gamma_{\mathrm{gap}} / 2}{(eV - E_{\mathrm{gap}})^2 + (\Gamma_{\mathrm{gap}} / 2)^2}$) with $A_{\mathrm{gap}}$ the amplitude, $\Gamma_{\mathrm{gap}}$ the gap full-width and $E_{\mathrm{gap}}$ its energy position. 
This combined function reproduces all experimental data very well (see purple dashed lines in Fig.\ref{Fig2}a). 
The additional gap is of $\approx$\,5\,meV width, albeit with site-dependent variations that will be discussed below.

This additional gap has not been reported at a temperature of 4-6\,K by Zhang et al. \cite{Zhang2018} or Zhao et al. \cite{Zhao2021_holes}. This discrepancy is easily resolved by raising the measurement temperature. Fig.\,\ref{Fig2}b shows the d$I$/d$V$ spectra at a single location measured with increasing temperature. To analyze the presence of the small gap, we again fit a combined function that includes a Fano lineshape and an additional gap to the spectra at all temperatures (Fig.\,\ref{Fig2}b). To better reveal the small gap, we plot in Fig.\,\ref{Fig2}c the spectra after subtracting the Fano lineshape. The resulting difference spectra clearly highlight the emergence of the small gap at low temperature and its suppression above 5\,K. 
The rapid suppression of the gap and its small energy scale indicate a low-energy phenomenon. Building on earlier predictions of a Kondo lattice \cite{Zhang2018, Zhao2021_holes}, we propose that this gap represents the previously elusive evidence of a hybridization gap, thereby supporting the identification of a Kondo lattice.

\begin{figure*}[t]
    \includegraphics[width=0.9\textwidth]{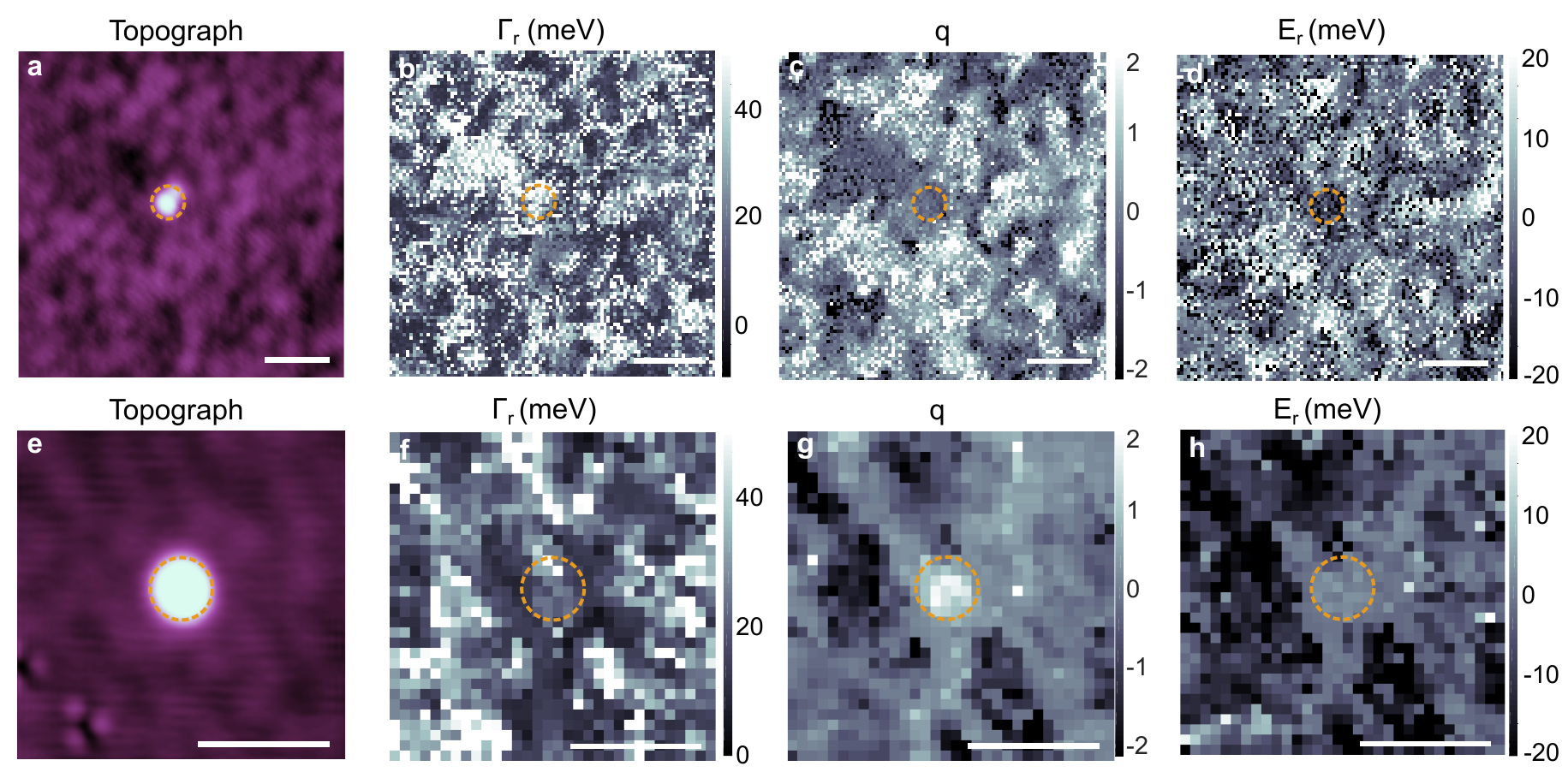}
	\caption{\label{Fig3} 
		Spatial variation of Fano parameters in presence of Fe and Mn atoms. (a,e) STM topographs around Fe (a) and Mn (e) atoms. The position of the atom is indicated by the orange circle. Scale bars are 2\,nm. Grids of d$I$/d$V$ spectra on these areas were fit with a Fano function combined with a gap as described in the text. The extracted Fano parameters are shown for Fe (b-d) and Mn (f-h) adatoms (resonance width $\Gamma_\mathrm{R}$, $q$ factor, and resonance energy $E_\mathrm{R}$). d$I$/d$V$ spectra recorded with b-d: V$_\mathrm{s}$ = 40 mV, I$_\mathrm{t}$ = 300 pA, V$_\mathrm{mod}$ = 0.5 mV), and f-h: V$_\mathrm{s}$ = 100 mV, I$_\mathrm{t}$ = 300 pA, V$_\mathrm{mod}$ = 1 mV).}
\end{figure*}

\begin{figure*}[t!]
    \includegraphics[width=0.9\textwidth]{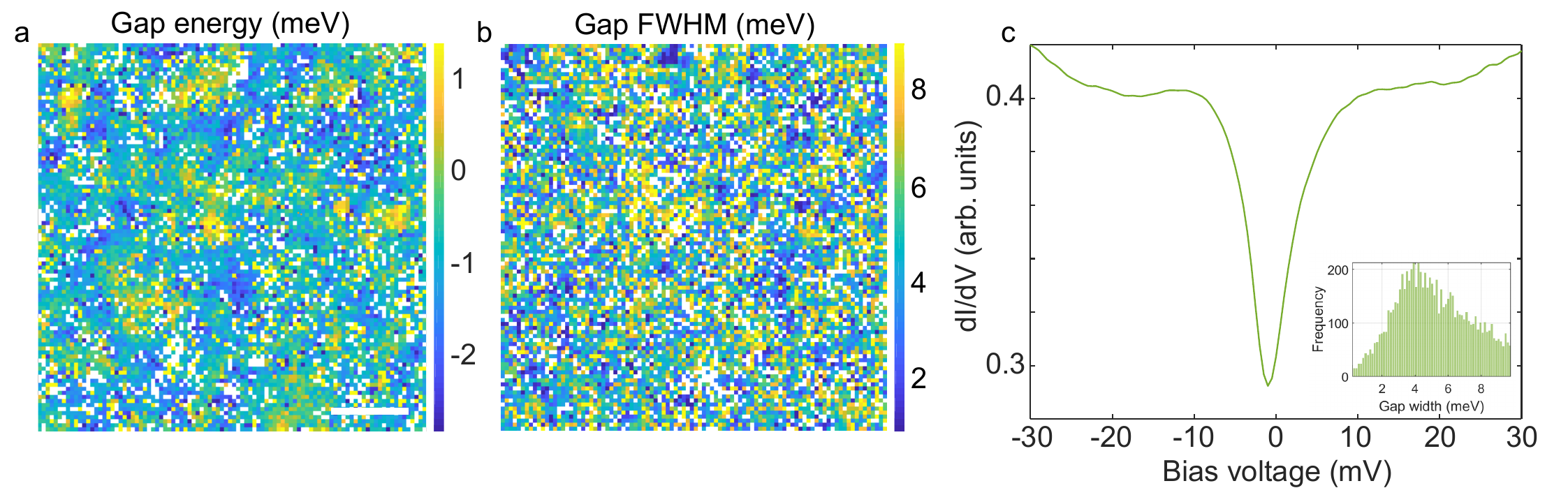}
	\caption{\label{Fig4} 
		Spatial variation of the hybridization gap. (a,b) Extracted gap energy and gap size from the grid of spectra shown in Fig.\,\ref{Fig3}a. The spectra are fit by a function that combines a Fano lineshape and a gap. (c) Plot of the average of all remaining gaps after subtraction of the Fano lineshape in the area of Fig.\,\ref{Fig3}a. The inset in Fig.\,\ref{Fig4}c shows the histogram of the gap FWHM values. The Fano contribution of the fit is subtracted from the spectra. Scale bars are 2\,nm.}
\end{figure*}

Next, we explore the potential Kondo resonance spatial variations and its response to single magnetic adatoms. To do so, we map out the spectral characteristics on the pristine Fe$_3$GeTe$_2$ surface and on single Fe (Fig.\,\ref{Fig3}a-d) and Mn (Fig.\,\ref{Fig3}e-h) atoms by recording d$I$/d$V$ spectra in a densely spaced grid covering the same area as the topography. We then apply our fit procedure using a Fano lineshape combined with an additional gap as described above. We first analyze the characteristics of the Fano contribution. We plot the extracted width $\Gamma_R$, asymmetry parameter $q$ and the position of the resonance $E_R$ in Fig.\,\ref{Fig3}b-d for Fe, and Fig.\,\ref{Fig3}f-h for Mn. 

First, we focus on the pristine background and note that all Kondo parameters exhibit spatial variations. Closer examination of the spatial distribution reveals no clear correlation of the topographic signal with any of the Kondo parameters, i.e., $\Gamma_R$, $q$, or $E_R$. This contrasts the observation by Zhao et al. who found that the asymmetry was changing across a Fe2 vacancy \cite{Zhao2021_holes}. In inspecting our topography, we note that most areas do no allow to identify isolated Fe2 vacancies, but a disordered modulation of the density of states, hinting at many different vacancies that may be located at different sample depths. This may preclude the observation of a direct correlation to specific topographic changes. However, we find a similar pattern, i.e., a direct correlation between the asymmetry parameter $q$ and the position of the Fano resonance $E_R$. This suggests that the local doping level determining the local electron-hole asymmetry is probed by the interfering tunneling paths. 

Surprisingly, the addition of magnetic Fe or Mn atoms does not lead to any noticeable variation of the Kondo correlation strength $\Gamma_\mathrm{R}$ beyond the variations on the pristine background (see Fig.\,\ref{Fig3}b,f). Furthermore, the Fe atom seems to be transparent to any of the other Kondo parameters, neither affecting the position of the Kondo resonance, nor the asymmetry. This is contrasted by the influence of the Mn atom, which leads to a significant change in the asymmetry $q$ of the resonance, suggesting that the different orbital occupation modifies the contributions of the co-tunneling paths to the interference.  

Next, we also investigate the spatial variation and response of the small gap to the presence of the Fe atom. Fig.\,\ref{Fig4}a shows the spatial variation of the gap energy and Fig.\,\ref{Fig4}b the corresponding width that has been extracted from the combined fit with the gap and Fano functions. The gap persists everywhere on the surface albeit with spatial variations (see inset in Fig.\,\ref{Fig4}b), reflecting the disorder already observed in topography. The average of the gap is shown in Fig.\,\ref{Fig4}c and amounts to $\approx$\,5\,meV. In comparing to the Fano parameters extracted in Fig.\,\ref{Fig3}b-d, we do not identify any spatial correlation. Surprisingly, we also do not find any impact of the Fe atom on the small hybridization gap. 

The absence of any noticeable effect on the correlated state of the underlying substrate by magnetic adsorbates suggests that the intralayer correlations are robust against external impurities. The exchange coupling of a weakly adsorbed atom on the terminating Te layer is, thus, insufficient to perturb the stronger correlations within the layers, which themselves are prone to variations imposed by Fe vacancies. 

In conclusion, we used scanning tunneling spectroscopy at 1.1\,K to investigate the low-energy excitation spectra on the surface of Fe$_3$GeTe$_2$. While we reproduced the previously observed Fano lineshape on the energy scale of $\approx$\,10\,meV, which has been interpreted as evidence of the Kondo effect, we additionally identified a small energy gap superimposed on this spectrum. Consistent with earlier predictions of heavy-fermion behavior, we attribute this gap to a hybridization gap arising from the long-range coherent interaction of Kondo impurities. These findings add to the debate regarding the presence of Kondo lattice behavior, which had appeared incompatible with the simultaneous ferromagnetic state. Further evidence for a robust long-range correlated state in the substrate was provided by the spatial mapping of the Kondo coupling parameters. The Kondo resonance as well as the hybridization gap persist all across the surface albeit with spatial variations that we ascribe to the inhomogeneous distribution of Fe vacancies in the substrate. While this suggest that magnetic properties can be tuned by the Fe content within the layer, magnetic adatoms on the surface do not significantly perturb the correlations within the layer, suggesting stronger intralayer interactions.
Our findings provide additional evidence of the complex magnetic properties of Fe$_3$GeTe$_2$ and underscore the need to consider the contributions of distinct Fe sites to fully explain its magnetic behavior.

	\section{Methods}
	
	The Fe$_3$GeTe$_2$ samples were purchased from HQgraphene. They were cleaved with a carbon tape at room temperature in the ultra-high vacuum chamber. The as-prepared crystals were transferred into a Joule-Thomson STM by Specs working at a base temperature of 1.1\,K. Fe and Mn atoms were deposited into the STM at a temperature below 10\,K.

	\section{Acknowledgements}
	
	We thank Fritz R\"ockenwagner for assistance in some of the experiments. We thank Nurit Avraham, Haim Beidenkopf and Ady Stern for sharing their data prior to publication and discussions. We thank Jose Lado for fruitful discussions. We acknowledge financial support by the Deutsche Forschungsgemeinschaft (DFG, German Research Foundation) through Project No. 328545488 (CRC 227, Project No. B05). CRV acknowledges funding from the European Union Horizon 2020 research and innovation programme under the Marie Skłodowska-Curie grant agreement No 844271).

\begin{thebibliography}{37}%
\makeatletter
\providecommand \@ifxundefined [1]{%
 \@ifx{#1\undefined}
}%
\providecommand \@ifnum [1]{%
 \ifnum #1\expandafter \@firstoftwo
 \else \expandafter \@secondoftwo
 \fi
}%
\providecommand \@ifx [1]{%
 \ifx #1\expandafter \@firstoftwo
 \else \expandafter \@secondoftwo
 \fi
}%
\providecommand \natexlab [1]{#1}%
\providecommand \enquote  [1]{``#1''}%
\providecommand \bibnamefont  [1]{#1}%
\providecommand \bibfnamefont [1]{#1}%
\providecommand \citenamefont [1]{#1}%
\providecommand \href@noop [0]{\@secondoftwo}%
\providecommand \href [0]{\begingroup \@sanitize@url \@href}%
\providecommand \@href[1]{\@@startlink{#1}\@@href}%
\providecommand \@@href[1]{\endgroup#1\@@endlink}%
\providecommand \@sanitize@url [0]{\catcode `\\12\catcode `\$12\catcode `\&12\catcode `\#12\catcode `\^12\catcode `\_12\catcode `\%12\relax}%
\providecommand \@@startlink[1]{}%
\providecommand \@@endlink[0]{}%
\providecommand \url  [0]{\begingroup\@sanitize@url \@url }%
\providecommand \@url [1]{\endgroup\@href {#1}{\urlprefix }}%
\providecommand \urlprefix  [0]{URL }%
\providecommand \Eprint [0]{\href }%
\providecommand \doibase [0]{https://doi.org/}%
\providecommand \selectlanguage [0]{\@gobble}%
\providecommand \bibinfo  [0]{\@secondoftwo}%
\providecommand \bibfield  [0]{\@secondoftwo}%
\providecommand \translation [1]{[#1]}%
\providecommand \BibitemOpen [0]{}%
\providecommand \bibitemStop [0]{}%
\providecommand \bibitemNoStop [0]{.\EOS\space}%
\providecommand \EOS [0]{\spacefactor3000\relax}%
\providecommand \BibitemShut  [1]{\csname bibitem#1\endcsname}%
\let\auto@bib@innerbib\@empty
\bibitem [{\citenamefont {Stewart}(1984)}]{Stewart1984}%
  \BibitemOpen
  \bibfield  {author} {\bibinfo {author} {\bibfnamefont {G.~R.}\ \bibnamefont {Stewart}},\ }\bibfield  {title} {\bibinfo {title} {Heavy-fermion systems},\ }\href {https://doi.org/10.1103/RevModPhys.56.755} {\bibfield  {journal} {\bibinfo  {journal} {Rev. Mod. Phys.}\ }\textbf {\bibinfo {volume} {56}},\ \bibinfo {pages} {755} (\bibinfo {year} {1984})}\BibitemShut {NoStop}%
\bibitem [{\citenamefont {Doniach}(1977)}]{Doniach1977}%
  \BibitemOpen
  \bibfield  {author} {\bibinfo {author} {\bibfnamefont {S.}~\bibnamefont {Doniach}},\ }\bibfield  {title} {\bibinfo {title} {{The Kondo lattice and weak antiferromagnetism}},\ }\href {https://doi.org/https://doi.org/10.1016/0378-4363(77)90190-5} {\bibfield  {journal} {\bibinfo  {journal} {Physica B+C}\ }\textbf {\bibinfo {volume} {91}},\ \bibinfo {pages} {231} (\bibinfo {year} {1977})}\BibitemShut {NoStop}%
\bibitem [{\citenamefont {Hewson}(1997)}]{Hewson1997}%
  \BibitemOpen
  \bibfield  {author} {\bibinfo {author} {\bibfnamefont {A.~C.}\ \bibnamefont {Hewson}},\ }\href@noop {} {\emph {\bibinfo {title} {The Kondo problem to heavy fermions}}}\ (\bibinfo  {publisher} {Cambridge University Press},\ \bibinfo {year} {1997})\BibitemShut {NoStop}%
\bibitem [{\citenamefont {Dzero}\ \emph {et~al.}(2016)\citenamefont {Dzero}, \citenamefont {Xia}, \citenamefont {Galitski},\ and\ \citenamefont {Coleman}}]{Dzero2016}%
  \BibitemOpen
  \bibfield  {author} {\bibinfo {author} {\bibfnamefont {M.}~\bibnamefont {Dzero}}, \bibinfo {author} {\bibfnamefont {J.}~\bibnamefont {Xia}}, \bibinfo {author} {\bibfnamefont {V.}~\bibnamefont {Galitski}},\ and\ \bibinfo {author} {\bibfnamefont {P.}~\bibnamefont {Coleman}},\ }\bibfield  {title} {\bibinfo {title} {{Topological Kondo Insulators}},\ }\href {https://doi.org/https://doi.org/10.1146/annurev-conmatphys-031214-014749} {\bibfield  {journal} {\bibinfo  {journal} {Ann. Rev. Cond. Matt. Phys.}\ }\textbf {\bibinfo {volume} {7}},\ \bibinfo {pages} {249} (\bibinfo {year} {2016})}\BibitemShut {NoStop}%
\bibitem [{\citenamefont {Schmidt}\ \emph {et~al.}(2010)\citenamefont {Schmidt}, \citenamefont {Hamidian}, \citenamefont {Wahl}, \citenamefont {Meier}, \citenamefont {Balatsky}, \citenamefont {Garrett}, \citenamefont {Williams}, \citenamefont {Luke},\ and\ \citenamefont {Davis}}]{Schmidt2010}%
  \BibitemOpen
  \bibfield  {author} {\bibinfo {author} {\bibfnamefont {A.~R.}\ \bibnamefont {Schmidt}}, \bibinfo {author} {\bibfnamefont {M.~H.}\ \bibnamefont {Hamidian}}, \bibinfo {author} {\bibfnamefont {P.}~\bibnamefont {Wahl}}, \bibinfo {author} {\bibfnamefont {F.}~\bibnamefont {Meier}}, \bibinfo {author} {\bibfnamefont {A.~V.}\ \bibnamefont {Balatsky}}, \bibinfo {author} {\bibfnamefont {J.~D.}\ \bibnamefont {Garrett}}, \bibinfo {author} {\bibfnamefont {T.~J.}\ \bibnamefont {Williams}}, \bibinfo {author} {\bibfnamefont {G.~M.}\ \bibnamefont {Luke}},\ and\ \bibinfo {author} {\bibfnamefont {J.~C.}\ \bibnamefont {Davis}},\ }\bibfield  {title} {\bibinfo {title} {{Imaging the Fano lattice to ‘hidden order’ transition in URu$_2$Si$_2$}},\ }\href {https://doi.org/10.1038/nature09073} {\bibfield  {journal} {\bibinfo  {journal} {Nature}\ }\textbf {\bibinfo {volume} {465}},\ \bibinfo {pages} {570–576} (\bibinfo {year} {2010})}\BibitemShut {NoStop}%
\bibitem [{\citenamefont {Ernst}\ \emph {et~al.}(2011)\citenamefont {Ernst}, \citenamefont {Kirchner}, \citenamefont {Krellner}, \citenamefont {Geibel}, \citenamefont {Zwicknagl}, \citenamefont {Steglich},\ and\ \citenamefont {Wirth}}]{Ernst2011}%
  \BibitemOpen
  \bibfield  {author} {\bibinfo {author} {\bibfnamefont {S.}~\bibnamefont {Ernst}}, \bibinfo {author} {\bibfnamefont {S.}~\bibnamefont {Kirchner}}, \bibinfo {author} {\bibfnamefont {C.}~\bibnamefont {Krellner}}, \bibinfo {author} {\bibfnamefont {C.}~\bibnamefont {Geibel}}, \bibinfo {author} {\bibfnamefont {G.}~\bibnamefont {Zwicknagl}}, \bibinfo {author} {\bibfnamefont {F.}~\bibnamefont {Steglich}},\ and\ \bibinfo {author} {\bibfnamefont {S.}~\bibnamefont {Wirth}},\ }\bibfield  {title} {\bibinfo {title} {{Emerging local Kondo screening and spatial coherence in the heavy-fermion metal YbRh$_2$Si$_2$}},\ }\href {https://doi.org/10.1038/nature10148} {\bibfield  {journal} {\bibinfo  {journal} {Nature}\ }\textbf {\bibinfo {volume} {474}},\ \bibinfo {pages} {362} (\bibinfo {year} {2011})}\BibitemShut {NoStop}%
\bibitem [{\citenamefont {Aynajian}\ \emph {et~al.}(2012)\citenamefont {Aynajian}, \citenamefont {da~Silva~Neto}, \citenamefont {Gyenis}, \citenamefont {Baumbach}, \citenamefont {Thompson}, \citenamefont {Fisk}, \citenamefont {Bauer},\ and\ \citenamefont {Yazdani}}]{Aynajian2012}%
  \BibitemOpen
  \bibfield  {author} {\bibinfo {author} {\bibfnamefont {P.}~\bibnamefont {Aynajian}}, \bibinfo {author} {\bibfnamefont {E.~H.}\ \bibnamefont {da~Silva~Neto}}, \bibinfo {author} {\bibfnamefont {A.}~\bibnamefont {Gyenis}}, \bibinfo {author} {\bibfnamefont {R.~E.}\ \bibnamefont {Baumbach}}, \bibinfo {author} {\bibfnamefont {J.~D.}\ \bibnamefont {Thompson}}, \bibinfo {author} {\bibfnamefont {Z.}~\bibnamefont {Fisk}}, \bibinfo {author} {\bibfnamefont {E.~D.}\ \bibnamefont {Bauer}},\ and\ \bibinfo {author} {\bibfnamefont {A.}~\bibnamefont {Yazdani}},\ }\bibfield  {title} {\bibinfo {title} {{Visualizing heavy fermions emerging in a quantum critical Kondo lattice}},\ }\href {https://doi.org/10.1038/nature11204} {\bibfield  {journal} {\bibinfo  {journal} {Nature}\ }\textbf {\bibinfo {volume} {486}},\ \bibinfo {pages} {201–206} (\bibinfo {year} {2012})}\BibitemShut {NoStop}%
\bibitem [{\citenamefont {Moro~Lagares}\ \emph {et~al.}(2019)\citenamefont {Moro~Lagares}, \citenamefont {Korytár}, \citenamefont {Piantek}, \citenamefont {Robles}, \citenamefont {Lorente}, \citenamefont {Pascual}, \citenamefont {Ibarra},\ and\ \citenamefont {Serrate}}]{Lagares2019}%
  \BibitemOpen
  \bibfield  {author} {\bibinfo {author} {\bibfnamefont {M.}~\bibnamefont {Moro~Lagares}}, \bibinfo {author} {\bibfnamefont {R.}~\bibnamefont {Korytár}}, \bibinfo {author} {\bibfnamefont {M.}~\bibnamefont {Piantek}}, \bibinfo {author} {\bibfnamefont {R.}~\bibnamefont {Robles}}, \bibinfo {author} {\bibfnamefont {N.}~\bibnamefont {Lorente}}, \bibinfo {author} {\bibfnamefont {J.}~\bibnamefont {Pascual}}, \bibinfo {author} {\bibfnamefont {M.}~\bibnamefont {Ibarra}},\ and\ \bibinfo {author} {\bibfnamefont {D.}~\bibnamefont {Serrate}},\ }\bibfield  {title} {\bibinfo {title} {{Real space manifestations of coherent screening in atomic scale Kondo lattices}},\ }\href {https://doi.org/10.1038/s41467-019-10103-5} {\bibfield  {journal} {\bibinfo  {journal} {Nature Commun.}\ }\textbf {\bibinfo {volume} {10}},\ \bibinfo {pages} {2211} (\bibinfo {year} {2019})}\BibitemShut {NoStop}%
\bibitem [{\citenamefont {Vaňo}\ \emph {et~al.}(2021)\citenamefont {Vaňo}, \citenamefont {Amini}, \citenamefont {Ganguli}, \citenamefont {Chen}, \citenamefont {Lado}, \citenamefont {Kezilebieke},\ and\ \citenamefont {Liljeroth}}]{Vano2021}%
  \BibitemOpen
  \bibfield  {author} {\bibinfo {author} {\bibfnamefont {V.}~\bibnamefont {Vaňo}}, \bibinfo {author} {\bibfnamefont {M.}~\bibnamefont {Amini}}, \bibinfo {author} {\bibfnamefont {S.~C.}\ \bibnamefont {Ganguli}}, \bibinfo {author} {\bibfnamefont {G.}~\bibnamefont {Chen}}, \bibinfo {author} {\bibfnamefont {J.~L.}\ \bibnamefont {Lado}}, \bibinfo {author} {\bibfnamefont {S.}~\bibnamefont {Kezilebieke}},\ and\ \bibinfo {author} {\bibfnamefont {P.}~\bibnamefont {Liljeroth}},\ }\bibfield  {title} {\bibinfo {title} {{Artificial heavy fermions in a van der Waals heterostructure}},\ }\href {https://doi.org/10.1038/s41586-021-04021-0} {\bibfield  {journal} {\bibinfo  {journal} {Nature}\ }\textbf {\bibinfo {volume} {599}},\ \bibinfo {pages} {582–586} (\bibinfo {year} {2021})}\BibitemShut {NoStop}%
\bibitem [{\citenamefont {Wan}\ \emph {et~al.}(2023)\citenamefont {Wan}, \citenamefont {Harsh}, \citenamefont {Meninno}, \citenamefont {Dreher}, \citenamefont {Sajan}, \citenamefont {Guo}, \citenamefont {Errea}, \citenamefont {de~Juan},\ and\ \citenamefont {Ugeda}}]{Wan2023}%
  \BibitemOpen
  \bibfield  {author} {\bibinfo {author} {\bibfnamefont {W.}~\bibnamefont {Wan}}, \bibinfo {author} {\bibfnamefont {R.}~\bibnamefont {Harsh}}, \bibinfo {author} {\bibfnamefont {A.}~\bibnamefont {Meninno}}, \bibinfo {author} {\bibfnamefont {P.}~\bibnamefont {Dreher}}, \bibinfo {author} {\bibfnamefont {S.}~\bibnamefont {Sajan}}, \bibinfo {author} {\bibfnamefont {H.}~\bibnamefont {Guo}}, \bibinfo {author} {\bibfnamefont {I.}~\bibnamefont {Errea}}, \bibinfo {author} {\bibfnamefont {F.}~\bibnamefont {de~Juan}},\ and\ \bibinfo {author} {\bibfnamefont {M.}~\bibnamefont {Ugeda}},\ }\bibfield  {title} {\bibinfo {title} {{Evidence for ground state coherence in a two-dimensional Kondo lattice}},\ }\href {https://doi.org/10.1038/s41467-023-42803-4} {\bibfield  {journal} {\bibinfo  {journal} {Nature Commun.}\ }\textbf {\bibinfo {volume} {14}},\ \bibinfo {pages} {7005} (\bibinfo {year} {2023})}\BibitemShut {NoStop}%
\bibitem [{\citenamefont {Novoselov}\ \emph {et~al.}(2016)\citenamefont {Novoselov}, \citenamefont {Mishchenko}, \citenamefont {Carvalho},\ and\ \citenamefont {Neto}}]{Novoselov2016}%
  \BibitemOpen
  \bibfield  {author} {\bibinfo {author} {\bibfnamefont {K.~S.}\ \bibnamefont {Novoselov}}, \bibinfo {author} {\bibfnamefont {A.}~\bibnamefont {Mishchenko}}, \bibinfo {author} {\bibfnamefont {A.}~\bibnamefont {Carvalho}},\ and\ \bibinfo {author} {\bibfnamefont {A.~H.~C.}\ \bibnamefont {Neto}},\ }\bibfield  {title} {\bibinfo {title} {{2D materials and van der Waals heterostructures}},\ }\href {https://doi.org/10.1126/science.aac9439} {\bibfield  {journal} {\bibinfo  {journal} {Science}\ }\textbf {\bibinfo {volume} {353}},\ \bibinfo {pages} {461} (\bibinfo {year} {2016})}\BibitemShut {NoStop}%
\bibitem [{\citenamefont {Gibertini}\ \emph {et~al.}(2019)\citenamefont {Gibertini}, \citenamefont {Koperski}, \citenamefont {Morpurgo},\ and\ \citenamefont {Novoselov}}]{Gibertini2019}%
  \BibitemOpen
  \bibfield  {author} {\bibinfo {author} {\bibfnamefont {M.}~\bibnamefont {Gibertini}}, \bibinfo {author} {\bibfnamefont {M.}~\bibnamefont {Koperski}}, \bibinfo {author} {\bibfnamefont {A.~F.}\ \bibnamefont {Morpurgo}},\ and\ \bibinfo {author} {\bibfnamefont {K.~S.}\ \bibnamefont {Novoselov}},\ }\bibfield  {title} {\bibinfo {title} {{Magnetic 2D materials and heterostructures}},\ }\href {https://doi.org/10.1038/s41565-019-0438-6} {\bibfield  {journal} {\bibinfo  {journal} {Nature Nanotechnol.}\ }\textbf {\bibinfo {volume} {14}},\ \bibinfo {pages} {408–419} (\bibinfo {year} {2019})}\BibitemShut {NoStop}%
\bibitem [{\citenamefont {Zhang}\ \emph {et~al.}(2018)\citenamefont {Zhang}, \citenamefont {Lu}, \citenamefont {Zhu}, \citenamefont {Tan}, \citenamefont {Feng}, \citenamefont {Liu}, \citenamefont {Zhang}, \citenamefont {Chen}, \citenamefont {Liu}, \citenamefont {Luo}, \citenamefont {Xie}, \citenamefont {Luo}, \citenamefont {Zhang},\ and\ \citenamefont {Lai}}]{Zhang2018}%
  \BibitemOpen
  \bibfield  {author} {\bibinfo {author} {\bibfnamefont {Y.}~\bibnamefont {Zhang}}, \bibinfo {author} {\bibfnamefont {H.}~\bibnamefont {Lu}}, \bibinfo {author} {\bibfnamefont {X.}~\bibnamefont {Zhu}}, \bibinfo {author} {\bibfnamefont {S.}~\bibnamefont {Tan}}, \bibinfo {author} {\bibfnamefont {W.}~\bibnamefont {Feng}}, \bibinfo {author} {\bibfnamefont {Q.}~\bibnamefont {Liu}}, \bibinfo {author} {\bibfnamefont {W.}~\bibnamefont {Zhang}}, \bibinfo {author} {\bibfnamefont {Q.}~\bibnamefont {Chen}}, \bibinfo {author} {\bibfnamefont {Y.}~\bibnamefont {Liu}}, \bibinfo {author} {\bibfnamefont {X.}~\bibnamefont {Luo}}, \bibinfo {author} {\bibfnamefont {D.}~\bibnamefont {Xie}}, \bibinfo {author} {\bibfnamefont {L.}~\bibnamefont {Luo}}, \bibinfo {author} {\bibfnamefont {Z.}~\bibnamefont {Zhang}},\ and\ \bibinfo {author} {\bibfnamefont {X.}~\bibnamefont {Lai}},\ }\bibfield  {title} {\bibinfo {title} {{Emergence of Kondo lattice behavior in a van der Waals itinerant ferromagnet, Fe$_3$GeTe$_2$}},\ }\href
  {https://doi.org/10.1126/sciadv.aao6791} {\bibfield  {journal} {\bibinfo  {journal} {Science Adv.}\ }\textbf {\bibinfo {volume} {4}},\ \bibinfo {pages} {eaao6791} (\bibinfo {year} {2018})}\BibitemShut {NoStop}%
\bibitem [{\citenamefont {Huang}\ \emph {et~al.}(2017)\citenamefont {Huang}, \citenamefont {Clark}, \citenamefont {Navarro-Moratalla}, \citenamefont {Klein}, \citenamefont {Cheng}, \citenamefont {Seyler}, \citenamefont {Zhong}, \citenamefont {Schmidgall}, \citenamefont {McGuire}, \citenamefont {Cobden}, \citenamefont {Yao}, \citenamefont {Xiao}, \citenamefont {Jarillo-Herrero},\ and\ \citenamefont {Xu}}]{Huang2017}%
  \BibitemOpen
  \bibfield  {author} {\bibinfo {author} {\bibfnamefont {B.}~\bibnamefont {Huang}}, \bibinfo {author} {\bibfnamefont {G.}~\bibnamefont {Clark}}, \bibinfo {author} {\bibfnamefont {E.}~\bibnamefont {Navarro-Moratalla}}, \bibinfo {author} {\bibfnamefont {D.~R.}\ \bibnamefont {Klein}}, \bibinfo {author} {\bibfnamefont {R.}~\bibnamefont {Cheng}}, \bibinfo {author} {\bibfnamefont {K.~L.}\ \bibnamefont {Seyler}}, \bibinfo {author} {\bibfnamefont {D.}~\bibnamefont {Zhong}}, \bibinfo {author} {\bibfnamefont {E.}~\bibnamefont {Schmidgall}}, \bibinfo {author} {\bibfnamefont {M.~A.}\ \bibnamefont {McGuire}}, \bibinfo {author} {\bibfnamefont {D.~H.}\ \bibnamefont {Cobden}}, \bibinfo {author} {\bibfnamefont {W.}~\bibnamefont {Yao}}, \bibinfo {author} {\bibfnamefont {D.}~\bibnamefont {Xiao}}, \bibinfo {author} {\bibfnamefont {P.}~\bibnamefont {Jarillo-Herrero}},\ and\ \bibinfo {author} {\bibfnamefont {X.}~\bibnamefont {Xu}},\ }\bibfield  {title} {\bibinfo {title} {{Layer-dependent ferromagnetism in a van der Waals crystal
  down to the monolayer limit}},\ }\href {https://doi.org/10.1038/nature22391} {\bibfield  {journal} {\bibinfo  {journal} {Nature}\ }\textbf {\bibinfo {volume} {546}},\ \bibinfo {pages} {270–273} (\bibinfo {year} {2017})}\BibitemShut {NoStop}%
\bibitem [{\citenamefont {Gong}\ \emph {et~al.}(2017)\citenamefont {Gong}, \citenamefont {Li}, \citenamefont {Li}, \citenamefont {Ji}, \citenamefont {Stern}, \citenamefont {Xia}, \citenamefont {Cao}, \citenamefont {Bao}, \citenamefont {Wang}, \citenamefont {Wang}, \citenamefont {Qiu}, \citenamefont {Cava}, \citenamefont {Louie}, \citenamefont {Xia},\ and\ \citenamefont {Zhang}}]{Gong2017}%
  \BibitemOpen
  \bibfield  {author} {\bibinfo {author} {\bibfnamefont {C.}~\bibnamefont {Gong}}, \bibinfo {author} {\bibfnamefont {L.}~\bibnamefont {Li}}, \bibinfo {author} {\bibfnamefont {Z.}~\bibnamefont {Li}}, \bibinfo {author} {\bibfnamefont {H.}~\bibnamefont {Ji}}, \bibinfo {author} {\bibfnamefont {A.}~\bibnamefont {Stern}}, \bibinfo {author} {\bibfnamefont {Y.}~\bibnamefont {Xia}}, \bibinfo {author} {\bibfnamefont {T.}~\bibnamefont {Cao}}, \bibinfo {author} {\bibfnamefont {W.}~\bibnamefont {Bao}}, \bibinfo {author} {\bibfnamefont {C.}~\bibnamefont {Wang}}, \bibinfo {author} {\bibfnamefont {Y.}~\bibnamefont {Wang}}, \bibinfo {author} {\bibfnamefont {Z.~Q.}\ \bibnamefont {Qiu}}, \bibinfo {author} {\bibfnamefont {R.~J.}\ \bibnamefont {Cava}}, \bibinfo {author} {\bibfnamefont {S.~G.}\ \bibnamefont {Louie}}, \bibinfo {author} {\bibfnamefont {J.}~\bibnamefont {Xia}},\ and\ \bibinfo {author} {\bibfnamefont {X.}~\bibnamefont {Zhang}},\ }\bibfield  {title} {\bibinfo {title} {{Discovery of intrinsic ferromagnetism in
  two-dimensional van der Waals crystals}},\ }\href {https://doi.org/10.1038/nature22060} {\bibfield  {journal} {\bibinfo  {journal} {Nature}\ }\textbf {\bibinfo {volume} {546}},\ \bibinfo {pages} {265–269} (\bibinfo {year} {2017})}\BibitemShut {NoStop}%
\bibitem [{\citenamefont {Deiseroth}\ \emph {et~al.}(2006)\citenamefont {Deiseroth}, \citenamefont {Aleksandrov}, \citenamefont {Reiner}, \citenamefont {Kienle},\ and\ \citenamefont {Kremer}}]{Deiseroth2006}%
  \BibitemOpen
  \bibfield  {author} {\bibinfo {author} {\bibfnamefont {H.-J.}\ \bibnamefont {Deiseroth}}, \bibinfo {author} {\bibfnamefont {K.}~\bibnamefont {Aleksandrov}}, \bibinfo {author} {\bibfnamefont {C.}~\bibnamefont {Reiner}}, \bibinfo {author} {\bibfnamefont {L.}~\bibnamefont {Kienle}},\ and\ \bibinfo {author} {\bibfnamefont {R.~K.}\ \bibnamefont {Kremer}},\ }\bibfield  {title} {\bibinfo {title} {{Fe$_3$GeTe$_2$ and Ni$_3$GeTe$_2$ – Two New Layered Transition-Metal Compounds: Crystal Structures, HRTEM Investigations, and Magnetic and Electrical Properties}},\ }\href {https://doi.org/https://doi.org/10.1002/ejic.200501020} {\bibfield  {journal} {\bibinfo  {journal} {Eur. J. Inorg. Chem.}\ }\textbf {\bibinfo {volume} {2006}},\ \bibinfo {pages} {1561} (\bibinfo {year} {2006})}\BibitemShut {NoStop}%
\bibitem [{\citenamefont {Chen}\ \emph {et~al.}(2013)\citenamefont {Chen}, \citenamefont {Yang}, \citenamefont {Wang}, \citenamefont {Imai}, \citenamefont {Ohta}, \citenamefont {Michioka}, \citenamefont {Yoshimura},\ and\ \citenamefont {Fang}}]{Chen2013}%
  \BibitemOpen
  \bibfield  {author} {\bibinfo {author} {\bibfnamefont {B.}~\bibnamefont {Chen}}, \bibinfo {author} {\bibfnamefont {J.}~\bibnamefont {Yang}}, \bibinfo {author} {\bibfnamefont {H.~D.}\ \bibnamefont {Wang}}, \bibinfo {author} {\bibfnamefont {M.}~\bibnamefont {Imai}}, \bibinfo {author} {\bibfnamefont {H.}~\bibnamefont {Ohta}}, \bibinfo {author} {\bibfnamefont {C.}~\bibnamefont {Michioka}}, \bibinfo {author} {\bibfnamefont {K.}~\bibnamefont {Yoshimura}},\ and\ \bibinfo {author} {\bibfnamefont {M.~H.}\ \bibnamefont {Fang}},\ }\bibfield  {title} {\bibinfo {title} {{Magnetic Properties of Layered Itinerant Electron Ferromagnet Fe$_3$GeTe$_2$}},\ }\href {https://doi.org/10.7566/JPSJ.82.124711} {\bibfield  {journal} {\bibinfo  {journal} {J. Phys. Soc. Jap.}\ }\textbf {\bibinfo {volume} {82}},\ \bibinfo {pages} {124711} (\bibinfo {year} {2013})}\BibitemShut {NoStop}%
\bibitem [{\citenamefont {May}\ \emph {et~al.}(2016)\citenamefont {May}, \citenamefont {Calder}, \citenamefont {Cantoni}, \citenamefont {Cao},\ and\ \citenamefont {McGuire}}]{May2016}%
  \BibitemOpen
  \bibfield  {author} {\bibinfo {author} {\bibfnamefont {A.~F.}\ \bibnamefont {May}}, \bibinfo {author} {\bibfnamefont {S.}~\bibnamefont {Calder}}, \bibinfo {author} {\bibfnamefont {C.}~\bibnamefont {Cantoni}}, \bibinfo {author} {\bibfnamefont {H.}~\bibnamefont {Cao}},\ and\ \bibinfo {author} {\bibfnamefont {M.~A.}\ \bibnamefont {McGuire}},\ }\bibfield  {title} {\bibinfo {title} {{Magnetic structure and phase stability of the van der Waals bonded ferromagnet ${\mathrm{Fe}}_{3\ensuremath{-}x}{\mathrm{GeTe}}_{2}$}},\ }\href {https://doi.org/10.1103/PhysRevB.93.014411} {\bibfield  {journal} {\bibinfo  {journal} {Phys. Rev. B}\ }\textbf {\bibinfo {volume} {93}},\ \bibinfo {pages} {014411} (\bibinfo {year} {2016})}\BibitemShut {NoStop}%
\bibitem [{\citenamefont {Fei}\ \emph {et~al.}(2018)\citenamefont {Fei}, \citenamefont {Huang}, \citenamefont {Malinowski}, \citenamefont {Wang}, \citenamefont {Song}, \citenamefont {Sanchez}, \citenamefont {Yao}, \citenamefont {Xiao}, \citenamefont {Zhu}, \citenamefont {May}, \citenamefont {Wu}, \citenamefont {Cobden}, \citenamefont {Chu},\ and\ \citenamefont {Xu}}]{Fei2018}%
  \BibitemOpen
  \bibfield  {author} {\bibinfo {author} {\bibfnamefont {Z.}~\bibnamefont {Fei}}, \bibinfo {author} {\bibfnamefont {B.}~\bibnamefont {Huang}}, \bibinfo {author} {\bibfnamefont {P.}~\bibnamefont {Malinowski}}, \bibinfo {author} {\bibfnamefont {W.}~\bibnamefont {Wang}}, \bibinfo {author} {\bibfnamefont {T.}~\bibnamefont {Song}}, \bibinfo {author} {\bibfnamefont {J.}~\bibnamefont {Sanchez}}, \bibinfo {author} {\bibfnamefont {W.}~\bibnamefont {Yao}}, \bibinfo {author} {\bibfnamefont {D.}~\bibnamefont {Xiao}}, \bibinfo {author} {\bibfnamefont {X.}~\bibnamefont {Zhu}}, \bibinfo {author} {\bibfnamefont {A.~F.}\ \bibnamefont {May}}, \bibinfo {author} {\bibfnamefont {W.}~\bibnamefont {Wu}}, \bibinfo {author} {\bibfnamefont {D.~H.}\ \bibnamefont {Cobden}}, \bibinfo {author} {\bibfnamefont {J.-H.}\ \bibnamefont {Chu}},\ and\ \bibinfo {author} {\bibfnamefont {X.}~\bibnamefont {Xu}},\ }\bibfield  {title} {\bibinfo {title} {{Two-dimensional itinerant ferromagnetism in atomically thin Fe$_3$GeTe$_2$}},\ }\href
  {https://doi.org/10.1038/s41563-018-0149-7} {\bibfield  {journal} {\bibinfo  {journal} {Nature Mater.}\ }\textbf {\bibinfo {volume} {17}},\ \bibinfo {pages} {778–782} (\bibinfo {year} {2018})}\BibitemShut {NoStop}%
\bibitem [{\citenamefont {Xu}\ \emph {et~al.}(2020)\citenamefont {Xu}, \citenamefont {Li}, \citenamefont {Duan}, \citenamefont {Zhang}, \citenamefont {Chen}, \citenamefont {Kang}, \citenamefont {Liang}, \citenamefont {Chen}, \citenamefont {Xia}, \citenamefont {Xu}, \citenamefont {Malinowski}, \citenamefont {Xu}, \citenamefont {Chu}, \citenamefont {Li}, \citenamefont {Guo}, \citenamefont {Liu}, \citenamefont {Yang},\ and\ \citenamefont {Chen}}]{Xu2020}%
  \BibitemOpen
  \bibfield  {author} {\bibinfo {author} {\bibfnamefont {X.}~\bibnamefont {Xu}}, \bibinfo {author} {\bibfnamefont {Y.~W.}\ \bibnamefont {Li}}, \bibinfo {author} {\bibfnamefont {S.~R.}\ \bibnamefont {Duan}}, \bibinfo {author} {\bibfnamefont {S.~L.}\ \bibnamefont {Zhang}}, \bibinfo {author} {\bibfnamefont {Y.~J.}\ \bibnamefont {Chen}}, \bibinfo {author} {\bibfnamefont {L.}~\bibnamefont {Kang}}, \bibinfo {author} {\bibfnamefont {A.~J.}\ \bibnamefont {Liang}}, \bibinfo {author} {\bibfnamefont {C.}~\bibnamefont {Chen}}, \bibinfo {author} {\bibfnamefont {W.}~\bibnamefont {Xia}}, \bibinfo {author} {\bibfnamefont {Y.}~\bibnamefont {Xu}}, \bibinfo {author} {\bibfnamefont {P.}~\bibnamefont {Malinowski}}, \bibinfo {author} {\bibfnamefont {X.~D.}\ \bibnamefont {Xu}}, \bibinfo {author} {\bibfnamefont {J.-H.}\ \bibnamefont {Chu}}, \bibinfo {author} {\bibfnamefont {G.}~\bibnamefont {Li}}, \bibinfo {author} {\bibfnamefont {Y.~F.}\ \bibnamefont {Guo}}, \bibinfo {author} {\bibfnamefont {Z.~K.}\ \bibnamefont {Liu}}, \bibinfo
  {author} {\bibfnamefont {L.~X.}\ \bibnamefont {Yang}},\ and\ \bibinfo {author} {\bibfnamefont {Y.~L.}\ \bibnamefont {Chen}},\ }\bibfield  {title} {\bibinfo {title} {{Signature for non-Stoner ferromagnetism in the van der Waals ferromagnet $\mathrm{F}{\mathrm{e}}_{3}\mathrm{GeT}{\mathrm{e}}_{2}$}},\ }\href {https://doi.org/10.1103/PhysRevB.101.201104} {\bibfield  {journal} {\bibinfo  {journal} {Phys. Rev. B}\ }\textbf {\bibinfo {volume} {101}},\ \bibinfo {pages} {201104} (\bibinfo {year} {2020})}\BibitemShut {NoStop}%
\bibitem [{\citenamefont {Yang}\ \emph {et~al.}(2022)\citenamefont {Yang}, \citenamefont {Bansal}, \citenamefont {Rüßmann}, \citenamefont {Hoffmann}, \citenamefont {Zhang}, \citenamefont {Go}, \citenamefont {Li}, \citenamefont {Haghighirad}, \citenamefont {Sen}, \citenamefont {Blügel}, \citenamefont {Le~Tacon}, \citenamefont {Mokrousov},\ and\ \citenamefont {Wulfhekel}}]{Yang2022}%
  \BibitemOpen
  \bibfield  {author} {\bibinfo {author} {\bibfnamefont {H.-H.}\ \bibnamefont {Yang}}, \bibinfo {author} {\bibfnamefont {N.}~\bibnamefont {Bansal}}, \bibinfo {author} {\bibfnamefont {P.}~\bibnamefont {Rüßmann}}, \bibinfo {author} {\bibfnamefont {M.}~\bibnamefont {Hoffmann}}, \bibinfo {author} {\bibfnamefont {L.}~\bibnamefont {Zhang}}, \bibinfo {author} {\bibfnamefont {D.}~\bibnamefont {Go}}, \bibinfo {author} {\bibfnamefont {Q.}~\bibnamefont {Li}}, \bibinfo {author} {\bibfnamefont {A.-A.}\ \bibnamefont {Haghighirad}}, \bibinfo {author} {\bibfnamefont {K.}~\bibnamefont {Sen}}, \bibinfo {author} {\bibfnamefont {S.}~\bibnamefont {Blügel}}, \bibinfo {author} {\bibfnamefont {M.}~\bibnamefont {Le~Tacon}}, \bibinfo {author} {\bibfnamefont {Y.}~\bibnamefont {Mokrousov}},\ and\ \bibinfo {author} {\bibfnamefont {W.}~\bibnamefont {Wulfhekel}},\ }\bibfield  {title} {\bibinfo {title} {{Magnetic domain walls of the van der Waals material Fe$_3$GeTe$_2$}},\ }\href {https://doi.org/10.1088/2053-1583/ac5d0e} {\bibfield
  {journal} {\bibinfo  {journal} {2D Materials}\ }\textbf {\bibinfo {volume} {9}},\ \bibinfo {pages} {025022} (\bibinfo {year} {2022})}\BibitemShut {NoStop}%
\bibitem [{\citenamefont {Zhao}\ \emph {et~al.}(2021)\citenamefont {Zhao}, \citenamefont {Chen}, \citenamefont {Xi}, \citenamefont {Zhao}, \citenamefont {Xu}, \citenamefont {Zhang}, \citenamefont {Cheng}, \citenamefont {Feng}, \citenamefont {Zhuang}, \citenamefont {Pan}, \citenamefont {Xu}, \citenamefont {Hao}, \citenamefont {Li}, \citenamefont {Zhou}, \citenamefont {Dou},\ and\ \citenamefont {Du}}]{Zhao2021_holes}%
  \BibitemOpen
  \bibfield  {author} {\bibinfo {author} {\bibfnamefont {M.}~\bibnamefont {Zhao}}, \bibinfo {author} {\bibfnamefont {B.-B.}\ \bibnamefont {Chen}}, \bibinfo {author} {\bibfnamefont {Y.}~\bibnamefont {Xi}}, \bibinfo {author} {\bibfnamefont {Y.}~\bibnamefont {Zhao}}, \bibinfo {author} {\bibfnamefont {H.}~\bibnamefont {Xu}}, \bibinfo {author} {\bibfnamefont {H.}~\bibnamefont {Zhang}}, \bibinfo {author} {\bibfnamefont {N.}~\bibnamefont {Cheng}}, \bibinfo {author} {\bibfnamefont {H.}~\bibnamefont {Feng}}, \bibinfo {author} {\bibfnamefont {J.}~\bibnamefont {Zhuang}}, \bibinfo {author} {\bibfnamefont {F.}~\bibnamefont {Pan}}, \bibinfo {author} {\bibfnamefont {X.}~\bibnamefont {Xu}}, \bibinfo {author} {\bibfnamefont {W.}~\bibnamefont {Hao}}, \bibinfo {author} {\bibfnamefont {W.}~\bibnamefont {Li}}, \bibinfo {author} {\bibfnamefont {S.}~\bibnamefont {Zhou}}, \bibinfo {author} {\bibfnamefont {S.~X.}\ \bibnamefont {Dou}},\ and\ \bibinfo {author} {\bibfnamefont {Y.}~\bibnamefont {Du}},\ }\bibfield  {title} {\bibinfo
  {title} {{Kondo Holes in the Two-Dimensional Itinerant Ising Ferromagnet Fe$_3$GeTe$_2$}},\ }\href {https://doi.org/10.1021/acs.nanolett.1c01661} {\bibfield  {journal} {\bibinfo  {journal} {Nano Lett.}\ }\textbf {\bibinfo {volume} {21}},\ \bibinfo {pages} {6117} (\bibinfo {year} {2021})}\BibitemShut {NoStop}%
\bibitem [{\citenamefont {Bao}\ \emph {et~al.}(2022)\citenamefont {Bao}, \citenamefont {Wang}, \citenamefont {Shangguan}, \citenamefont {Cai}, \citenamefont {Dong}, \citenamefont {Huang}, \citenamefont {Si}, \citenamefont {Ma}, \citenamefont {Kajimoto}, \citenamefont {Ikeuchi}, \citenamefont {Yano}, \citenamefont {Yu}, \citenamefont {Wan}, \citenamefont {Li},\ and\ \citenamefont {Wen}}]{Bao2022}%
  \BibitemOpen
  \bibfield  {author} {\bibinfo {author} {\bibfnamefont {S.}~\bibnamefont {Bao}}, \bibinfo {author} {\bibfnamefont {W.}~\bibnamefont {Wang}}, \bibinfo {author} {\bibfnamefont {Y.}~\bibnamefont {Shangguan}}, \bibinfo {author} {\bibfnamefont {Z.}~\bibnamefont {Cai}}, \bibinfo {author} {\bibfnamefont {Z.-Y.}\ \bibnamefont {Dong}}, \bibinfo {author} {\bibfnamefont {Z.}~\bibnamefont {Huang}}, \bibinfo {author} {\bibfnamefont {W.}~\bibnamefont {Si}}, \bibinfo {author} {\bibfnamefont {Z.}~\bibnamefont {Ma}}, \bibinfo {author} {\bibfnamefont {R.}~\bibnamefont {Kajimoto}}, \bibinfo {author} {\bibfnamefont {K.}~\bibnamefont {Ikeuchi}}, \bibinfo {author} {\bibfnamefont {S.-i.}\ \bibnamefont {Yano}}, \bibinfo {author} {\bibfnamefont {S.-L.}\ \bibnamefont {Yu}}, \bibinfo {author} {\bibfnamefont {X.}~\bibnamefont {Wan}}, \bibinfo {author} {\bibfnamefont {J.-X.}\ \bibnamefont {Li}},\ and\ \bibinfo {author} {\bibfnamefont {J.}~\bibnamefont {Wen}},\ }\bibfield  {title} {\bibinfo {title} {{Neutron Spectroscopy Evidence on the
  Dual Nature of Magnetic Excitations in a van der Waals Metallic Ferromagnet ${\mathrm{Fe}}_{2.72}{\mathrm{GeTe}}_{2}$}},\ }\href {https://doi.org/10.1103/PhysRevX.12.011022} {\bibfield  {journal} {\bibinfo  {journal} {Phys. Rev. X}\ }\textbf {\bibinfo {volume} {12}},\ \bibinfo {pages} {011022} (\bibinfo {year} {2022})}\BibitemShut {NoStop}%
\bibitem [{\citenamefont {Bai}\ \emph {et~al.}(2022)\citenamefont {Bai}, \citenamefont {Lechermann}, \citenamefont {Liu}, \citenamefont {Cheng}, \citenamefont {Kolesnikov}, \citenamefont {Ye}, \citenamefont {Williams}, \citenamefont {Chi}, \citenamefont {Hong}, \citenamefont {Granroth}, \citenamefont {May},\ and\ \citenamefont {Calder}}]{Bai2022}%
  \BibitemOpen
  \bibfield  {author} {\bibinfo {author} {\bibfnamefont {X.}~\bibnamefont {Bai}}, \bibinfo {author} {\bibfnamefont {F.}~\bibnamefont {Lechermann}}, \bibinfo {author} {\bibfnamefont {Y.}~\bibnamefont {Liu}}, \bibinfo {author} {\bibfnamefont {Y.}~\bibnamefont {Cheng}}, \bibinfo {author} {\bibfnamefont {A.~I.}\ \bibnamefont {Kolesnikov}}, \bibinfo {author} {\bibfnamefont {F.}~\bibnamefont {Ye}}, \bibinfo {author} {\bibfnamefont {T.~J.}\ \bibnamefont {Williams}}, \bibinfo {author} {\bibfnamefont {S.}~\bibnamefont {Chi}}, \bibinfo {author} {\bibfnamefont {T.}~\bibnamefont {Hong}}, \bibinfo {author} {\bibfnamefont {G.~E.}\ \bibnamefont {Granroth}}, \bibinfo {author} {\bibfnamefont {A.~F.}\ \bibnamefont {May}},\ and\ \bibinfo {author} {\bibfnamefont {S.}~\bibnamefont {Calder}},\ }\bibfield  {title} {\bibinfo {title} {{Antiferromagnetic fluctuations and orbital-selective Mott transition in the van der Waals ferromagnet Fe$_{3-x}$GeTe$_2$}},\ }\href {https://doi.org/10.1103/PhysRevB.106.L180409} {\bibfield  {journal}
  {\bibinfo  {journal} {Phys. Rev. B}\ }\textbf {\bibinfo {volume} {106}},\ \bibinfo {pages} {L180409} (\bibinfo {year} {2022})}\BibitemShut {NoStop}%
\bibitem [{\citenamefont {Zhu}\ \emph {et~al.}(2016)\citenamefont {Zhu}, \citenamefont {Janoschek}, \citenamefont {Chaves}, \citenamefont {Cezar}, \citenamefont {Durakiewicz}, \citenamefont {Ronning}, \citenamefont {Sassa}, \citenamefont {Mansson}, \citenamefont {Scott}, \citenamefont {Wakeham}, \citenamefont {Bauer},\ and\ \citenamefont {Thompson}}]{Zhu2016}%
  \BibitemOpen
  \bibfield  {author} {\bibinfo {author} {\bibfnamefont {J.-X.}\ \bibnamefont {Zhu}}, \bibinfo {author} {\bibfnamefont {M.}~\bibnamefont {Janoschek}}, \bibinfo {author} {\bibfnamefont {D.~S.}\ \bibnamefont {Chaves}}, \bibinfo {author} {\bibfnamefont {J.~C.}\ \bibnamefont {Cezar}}, \bibinfo {author} {\bibfnamefont {T.}~\bibnamefont {Durakiewicz}}, \bibinfo {author} {\bibfnamefont {F.}~\bibnamefont {Ronning}}, \bibinfo {author} {\bibfnamefont {Y.}~\bibnamefont {Sassa}}, \bibinfo {author} {\bibfnamefont {M.}~\bibnamefont {Mansson}}, \bibinfo {author} {\bibfnamefont {B.~L.}\ \bibnamefont {Scott}}, \bibinfo {author} {\bibfnamefont {N.}~\bibnamefont {Wakeham}}, \bibinfo {author} {\bibfnamefont {E.~D.}\ \bibnamefont {Bauer}},\ and\ \bibinfo {author} {\bibfnamefont {J.~D.}\ \bibnamefont {Thompson}},\ }\bibfield  {title} {\bibinfo {title} {{Electronic correlation and magnetism in the ferromagnetic metal ${\mathrm{Fe}}_{3}{\mathrm{GeTe}}_{2}$}},\ }\href {https://doi.org/10.1103/PhysRevB.93.144404} {\bibfield  {journal}
  {\bibinfo  {journal} {Phys. Rev. B}\ }\textbf {\bibinfo {volume} {93}},\ \bibinfo {pages} {144404} (\bibinfo {year} {2016})}\BibitemShut {NoStop}%
\bibitem [{\citenamefont {Corasaniti}\ \emph {et~al.}(2020)\citenamefont {Corasaniti}, \citenamefont {Yang}, \citenamefont {Sen}, \citenamefont {Willa}, \citenamefont {Merz}, \citenamefont {Haghighirad}, \citenamefont {Le~Tacon},\ and\ \citenamefont {Degiorgi}}]{Corasantini2020}%
  \BibitemOpen
  \bibfield  {author} {\bibinfo {author} {\bibfnamefont {M.}~\bibnamefont {Corasaniti}}, \bibinfo {author} {\bibfnamefont {R.}~\bibnamefont {Yang}}, \bibinfo {author} {\bibfnamefont {K.}~\bibnamefont {Sen}}, \bibinfo {author} {\bibfnamefont {K.}~\bibnamefont {Willa}}, \bibinfo {author} {\bibfnamefont {M.}~\bibnamefont {Merz}}, \bibinfo {author} {\bibfnamefont {A.~A.}\ \bibnamefont {Haghighirad}}, \bibinfo {author} {\bibfnamefont {M.}~\bibnamefont {Le~Tacon}},\ and\ \bibinfo {author} {\bibfnamefont {L.}~\bibnamefont {Degiorgi}},\ }\bibfield  {title} {\bibinfo {title} {{Electronic correlations in the van der Waals ferromagnet Fe$_3$GeTe$_2$ revealed by its charge dynamics}},\ }\href {https://doi.org/10.1103/PhysRevB.102.161109} {\bibfield  {journal} {\bibinfo  {journal} {Phys. Rev. B}\ }\textbf {\bibinfo {volume} {102}},\ \bibinfo {pages} {161109} (\bibinfo {year} {2020})}\BibitemShut {NoStop}%
\bibitem [{\citenamefont {Kim}\ \emph {et~al.}(2022)\citenamefont {Kim}, \citenamefont {Ryee},\ and\ \citenamefont {Han}}]{Kim2022}%
  \BibitemOpen
  \bibfield  {author} {\bibinfo {author} {\bibfnamefont {T.~J.}\ \bibnamefont {Kim}}, \bibinfo {author} {\bibfnamefont {S.}~\bibnamefont {Ryee}},\ and\ \bibinfo {author} {\bibfnamefont {M.~J.}\ \bibnamefont {Han}},\ }\bibfield  {title} {\bibinfo {title} {{Fe$_3$GeTe$_2$: a site-differentiated Hund metal}},\ }\href {http://dx.doi.org/10.1038/s41524-022-00937-x} {\bibfield  {journal} {\bibinfo  {journal} {npj Computational Mater.}\ }\textbf {\bibinfo {volume} {8}} (\bibinfo {year} {2022})}\BibitemShut {NoStop}%
\bibitem [{\citenamefont {Xu}\ \emph {et~al.}(2024)\citenamefont {Xu}, \citenamefont {Wang}, \citenamefont {Jin}, \citenamefont {Liu}, \citenamefont {Liu}, \citenamefont {Song},\ and\ \citenamefont {Tian}}]{Xu2024}%
  \BibitemOpen
  \bibfield  {author} {\bibinfo {author} {\bibfnamefont {Y.}~\bibnamefont {Xu}}, \bibinfo {author} {\bibfnamefont {Y.-C.}\ \bibnamefont {Wang}}, \bibinfo {author} {\bibfnamefont {X.}~\bibnamefont {Jin}}, \bibinfo {author} {\bibfnamefont {H.}~\bibnamefont {Liu}}, \bibinfo {author} {\bibfnamefont {Y.}~\bibnamefont {Liu}}, \bibinfo {author} {\bibfnamefont {H.}~\bibnamefont {Song}},\ and\ \bibinfo {author} {\bibfnamefont {F.}~\bibnamefont {Tian}},\ }\bibfield  {title} {\bibinfo {title} {{Mechanism of magnetic phase transition in correlated magnetic metal: insight into itinerant ferromagnet Fe$_{3-\delta}$GeTe$_2$}},\ }\href {http://dx.doi.org/10.1038/s42005-024-01874-5} {\bibfield  {journal} {\bibinfo  {journal} {Commun. Phys.}\ }\textbf {\bibinfo {volume} {7}} (\bibinfo {year} {2024})}\BibitemShut {NoStop}%
\bibitem [{\citenamefont {Bansal}\ \emph {et~al.}(2023)\citenamefont {Bansal}, \citenamefont {Li}, \citenamefont {Nufer}, \citenamefont {Zhang}, \citenamefont {Haghighirad}, \citenamefont {Mokrousov},\ and\ \citenamefont {Wulfhekel}}]{Bansal2023}%
  \BibitemOpen
  \bibfield  {author} {\bibinfo {author} {\bibfnamefont {N.}~\bibnamefont {Bansal}}, \bibinfo {author} {\bibfnamefont {Q.}~\bibnamefont {Li}}, \bibinfo {author} {\bibfnamefont {P.}~\bibnamefont {Nufer}}, \bibinfo {author} {\bibfnamefont {L.}~\bibnamefont {Zhang}}, \bibinfo {author} {\bibfnamefont {A.-A.}\ \bibnamefont {Haghighirad}}, \bibinfo {author} {\bibfnamefont {Y.}~\bibnamefont {Mokrousov}},\ and\ \bibinfo {author} {\bibfnamefont {W.}~\bibnamefont {Wulfhekel}},\ }\href {https://arxiv.org/abs/2308.10774} {\bibinfo {title} {{Magnon-Phonon coupling in Fe$_3$GeTe$_2$}}} (\bibinfo {year} {2023}),\ \Eprint {https://arxiv.org/abs/2308.10774} {arXiv:2308.10774 [cond-mat.mtrl-sci]} \BibitemShut {NoStop}%
\bibitem [{\citenamefont {Mathimalar}\ \emph {et~al.}(2025)\citenamefont {Mathimalar}, \citenamefont {Gupta}, \citenamefont {Roet}, \citenamefont {Galeski}, \citenamefont {Wawrzynczak}, \citenamefont {Garcia-Diez}, \citenamefont {Robredo}, \citenamefont {Vir}, \citenamefont {Kumar}, \citenamefont {Schnelle}, \citenamefont {von Arx}, \citenamefont {Küspert}, \citenamefont {Wang}, \citenamefont {Chang}, \citenamefont {Sassa}, \citenamefont {Stern}, \citenamefont {von Oppen}, \citenamefont {Vergniory}, \citenamefont {Felser}, \citenamefont {Gooth}, \citenamefont {Avraham},\ and\ \citenamefont {Beidenkopf}}]{Mathimalar2025}%
  \BibitemOpen
  \bibfield  {author} {\bibinfo {author} {\bibfnamefont {S.}~\bibnamefont {Mathimalar}}, \bibinfo {author} {\bibfnamefont {A.}~\bibnamefont {Gupta}}, \bibinfo {author} {\bibfnamefont {Y.}~\bibnamefont {Roet}}, \bibinfo {author} {\bibfnamefont {S.}~\bibnamefont {Galeski}}, \bibinfo {author} {\bibfnamefont {R.}~\bibnamefont {Wawrzynczak}}, \bibinfo {author} {\bibfnamefont {M.}~\bibnamefont {Garcia-Diez}}, \bibinfo {author} {\bibfnamefont {I.}~\bibnamefont {Robredo}}, \bibinfo {author} {\bibfnamefont {P.}~\bibnamefont {Vir}}, \bibinfo {author} {\bibfnamefont {N.}~\bibnamefont {Kumar}}, \bibinfo {author} {\bibfnamefont {W.}~\bibnamefont {Schnelle}}, \bibinfo {author} {\bibfnamefont {K.}~\bibnamefont {von Arx}}, \bibinfo {author} {\bibfnamefont {J.}~\bibnamefont {Küspert}}, \bibinfo {author} {\bibfnamefont {Q.}~\bibnamefont {Wang}}, \bibinfo {author} {\bibfnamefont {J.}~\bibnamefont {Chang}}, \bibinfo {author} {\bibfnamefont {Y.}~\bibnamefont {Sassa}}, \bibinfo {author} {\bibfnamefont {A.}~\bibnamefont {Stern}},
  \bibinfo {author} {\bibfnamefont {F.}~\bibnamefont {von Oppen}}, \bibinfo {author} {\bibfnamefont {M.~G.}\ \bibnamefont {Vergniory}}, \bibinfo {author} {\bibfnamefont {C.}~\bibnamefont {Felser}}, \bibinfo {author} {\bibfnamefont {J.}~\bibnamefont {Gooth}}, \bibinfo {author} {\bibfnamefont {N.}~\bibnamefont {Avraham}},\ and\ \bibinfo {author} {\bibfnamefont {H.}~\bibnamefont {Beidenkopf}},\ }\href {https://arxiv.org/abs/2503.04367} {\bibinfo {title} {Concurrent multifractality and anomalous hall response in the nodal line semimetal fe$_3$gete$_2$ near localization}} (\bibinfo {year} {2025}),\ \Eprint {https://arxiv.org/abs/2503.04367} {arXiv:2503.04367 [cond-mat.str-el]} \BibitemShut {NoStop}%
\bibitem [{\citenamefont {Hamidian}\ \emph {et~al.}(2011)\citenamefont {Hamidian}, \citenamefont {Schmidt}, \citenamefont {Firmo}, \citenamefont {Allan}, \citenamefont {Bradley}, \citenamefont {Garrett}, \citenamefont {Williams}, \citenamefont {Luke}, \citenamefont {Dubi}, \citenamefont {Balatsky},\ and\ \citenamefont {Davis}}]{Hamidian2011}%
  \BibitemOpen
  \bibfield  {author} {\bibinfo {author} {\bibfnamefont {M.~H.}\ \bibnamefont {Hamidian}}, \bibinfo {author} {\bibfnamefont {A.~R.}\ \bibnamefont {Schmidt}}, \bibinfo {author} {\bibfnamefont {I.~A.}\ \bibnamefont {Firmo}}, \bibinfo {author} {\bibfnamefont {M.~P.}\ \bibnamefont {Allan}}, \bibinfo {author} {\bibfnamefont {P.}~\bibnamefont {Bradley}}, \bibinfo {author} {\bibfnamefont {J.~D.}\ \bibnamefont {Garrett}}, \bibinfo {author} {\bibfnamefont {T.~J.}\ \bibnamefont {Williams}}, \bibinfo {author} {\bibfnamefont {G.~M.}\ \bibnamefont {Luke}}, \bibinfo {author} {\bibfnamefont {Y.}~\bibnamefont {Dubi}}, \bibinfo {author} {\bibfnamefont {A.~V.}\ \bibnamefont {Balatsky}},\ and\ \bibinfo {author} {\bibfnamefont {J.~C.}\ \bibnamefont {Davis}},\ }\bibfield  {title} {\bibinfo {title} {{How Kondo-holes create intense nanoscale heavy-fermion hybridization disorder}},\ }\href {https://doi.org/10.1073/pnas.1115027108} {\bibfield  {journal} {\bibinfo  {journal} {Proc. Nat. Acad. Sci.}\ }\textbf {\bibinfo {volume}
  {108}},\ \bibinfo {pages} {18233} (\bibinfo {year} {2011})}\BibitemShut {NoStop}%
\bibitem [{\citenamefont {Horcas}\ \emph {et~al.}(2007)\citenamefont {Horcas}, \citenamefont {Fernández}, \citenamefont {Gómez-Rodríguez}, \citenamefont {Colchero}, \citenamefont {Gómez-Herrero},\ and\ \citenamefont {Baro}}]{wsxm}%
  \BibitemOpen
  \bibfield  {author} {\bibinfo {author} {\bibfnamefont {I.}~\bibnamefont {Horcas}}, \bibinfo {author} {\bibfnamefont {R.}~\bibnamefont {Fernández}}, \bibinfo {author} {\bibfnamefont {J.~M.}\ \bibnamefont {Gómez-Rodríguez}}, \bibinfo {author} {\bibfnamefont {J.}~\bibnamefont {Colchero}}, \bibinfo {author} {\bibfnamefont {J.}~\bibnamefont {Gómez-Herrero}},\ and\ \bibinfo {author} {\bibfnamefont {A.~M.}\ \bibnamefont {Baro}},\ }\bibfield  {title} {\bibinfo {title} {Wsxm: A software for scanning probe microscopy and a tool for nanotechnology},\ }\href {https://doi.org/10.1063/1.2432410} {\bibfield  {journal} {\bibinfo  {journal} {Rev. Sci. Instr.}\ }\textbf {\bibinfo {volume} {78}},\ \bibinfo {pages} {013705} (\bibinfo {year} {2007})}\BibitemShut {NoStop}%
\bibitem [{\citenamefont {Ruby}(2016)}]{Spectrafox}%
  \BibitemOpen
  \bibfield  {author} {\bibinfo {author} {\bibfnamefont {M.}~\bibnamefont {Ruby}},\ }\bibfield  {title} {\bibinfo {title} {Spectrafox: A free open-source data management and analysis tool for scanning probe microscopy and spectroscopy},\ }\href {https://doi.org/https://doi.org/10.1016/j.softx.2016.04.001} {\bibfield  {journal} {\bibinfo  {journal} {SoftwareX}\ }\textbf {\bibinfo {volume} {5}},\ \bibinfo {pages} {31} (\bibinfo {year} {2016})}\BibitemShut {NoStop}%
\bibitem [{\citenamefont {Nguyen}\ \emph {et~al.}(2018)\citenamefont {Nguyen}, \citenamefont {Lee}, \citenamefont {Berlijn}, \citenamefont {Zou}, \citenamefont {Hus}, \citenamefont {Park}, \citenamefont {Gai}, \citenamefont {Lee},\ and\ \citenamefont {Li}}]{Nguyen2018}%
  \BibitemOpen
  \bibfield  {author} {\bibinfo {author} {\bibfnamefont {G.~D.}\ \bibnamefont {Nguyen}}, \bibinfo {author} {\bibfnamefont {J.}~\bibnamefont {Lee}}, \bibinfo {author} {\bibfnamefont {T.}~\bibnamefont {Berlijn}}, \bibinfo {author} {\bibfnamefont {Q.}~\bibnamefont {Zou}}, \bibinfo {author} {\bibfnamefont {S.~M.}\ \bibnamefont {Hus}}, \bibinfo {author} {\bibfnamefont {J.}~\bibnamefont {Park}}, \bibinfo {author} {\bibfnamefont {Z.}~\bibnamefont {Gai}}, \bibinfo {author} {\bibfnamefont {C.}~\bibnamefont {Lee}},\ and\ \bibinfo {author} {\bibfnamefont {A.-P.}\ \bibnamefont {Li}},\ }\bibfield  {title} {\bibinfo {title} {Visualization and manipulation of magnetic domains in the quasi-two-dimensional material $\mathrm{F}{\mathrm{e}}_{3}\mathrm{GeT}{\mathrm{e}}_{2}$},\ }\href {https://doi.org/10.1103/PhysRevB.97.014425} {\bibfield  {journal} {\bibinfo  {journal} {Phys. Rev. B}\ }\textbf {\bibinfo {volume} {97}},\ \bibinfo {pages} {014425} (\bibinfo {year} {2018})}\BibitemShut {NoStop}%
\bibitem [{\citenamefont {Trainer}\ \emph {et~al.}(2022)\citenamefont {Trainer}, \citenamefont {Armitage}, \citenamefont {Lane}, \citenamefont {Rhodes}, \citenamefont {Chan}, \citenamefont {Benedi\ifmmode \check{c}\else \v{c}\fi{}i\ifmmode~\check{c}\else \v{c}\fi{}}, \citenamefont {Rodriguez-Rivera}, \citenamefont {Fabelo}, \citenamefont {Stock},\ and\ \citenamefont {Wahl}}]{Trainer2022}%
  \BibitemOpen
  \bibfield  {author} {\bibinfo {author} {\bibfnamefont {C.}~\bibnamefont {Trainer}}, \bibinfo {author} {\bibfnamefont {O.~R.}\ \bibnamefont {Armitage}}, \bibinfo {author} {\bibfnamefont {H.}~\bibnamefont {Lane}}, \bibinfo {author} {\bibfnamefont {L.~C.}\ \bibnamefont {Rhodes}}, \bibinfo {author} {\bibfnamefont {E.}~\bibnamefont {Chan}}, \bibinfo {author} {\bibfnamefont {I.}~\bibnamefont {Benedi\ifmmode \check{c}\else \v{c}\fi{}i\ifmmode~\check{c}\else \v{c}\fi{}}}, \bibinfo {author} {\bibfnamefont {J.~A.}\ \bibnamefont {Rodriguez-Rivera}}, \bibinfo {author} {\bibfnamefont {O.}~\bibnamefont {Fabelo}}, \bibinfo {author} {\bibfnamefont {C.}~\bibnamefont {Stock}},\ and\ \bibinfo {author} {\bibfnamefont {P.}~\bibnamefont {Wahl}},\ }\bibfield  {title} {\bibinfo {title} {{Relating spin-polarized STM imaging and inelastic neutron scattering in the van der Waals ferromagnet ${\mathrm{Fe}}_{3}{\mathrm{GeTe}}_{2}$}},\ }\href {https://doi.org/10.1103/PhysRevB.106.L081405} {\bibfield  {journal} {\bibinfo  {journal}
  {Phys. Rev. B}\ }\textbf {\bibinfo {volume} {106}},\ \bibinfo {pages} {L081405} (\bibinfo {year} {2022})}\BibitemShut {NoStop}%
\bibitem [{\citenamefont {Yang}(2009)}]{Yang2009}%
  \BibitemOpen
  \bibfield  {author} {\bibinfo {author} {\bibfnamefont {Y.-f.}\ \bibnamefont {Yang}},\ }\bibfield  {title} {\bibinfo {title} {Fano effect in the point contact spectroscopy of heavy-electron materials},\ }\href {https://doi.org/10.1103/PhysRevB.79.241107} {\bibfield  {journal} {\bibinfo  {journal} {Phys. Rev. B}\ }\textbf {\bibinfo {volume} {79}},\ \bibinfo {pages} {241107} (\bibinfo {year} {2009})}\BibitemShut {NoStop}%
\bibitem [{\citenamefont {Figgins}\ and\ \citenamefont {Morr}(2011)}]{Figgins2011}%
  \BibitemOpen
  \bibfield  {author} {\bibinfo {author} {\bibfnamefont {J.}~\bibnamefont {Figgins}}\ and\ \bibinfo {author} {\bibfnamefont {D.~K.}\ \bibnamefont {Morr}},\ }\bibfield  {title} {\bibinfo {title} {{Defects in Heavy-Fermion Materials: Unveiling Strong Correlations in Real Space}},\ }\href {https://doi.org/10.1103/PhysRevLett.107.066401} {\bibfield  {journal} {\bibinfo  {journal} {Phys. Rev. Lett.}\ }\textbf {\bibinfo {volume} {107}},\ \bibinfo {pages} {066401} (\bibinfo {year} {2011})}\BibitemShut {NoStop}%
\end{thebibliography}

    %

	%
\end{document}